\documentstyle[12pt]{article}
\begin{document}
\setlength{\baselineskip}{0.30in}

\newcommand{\be}{\begin{equation}}
\newcommand{\ee}{\end{equation}}
\newcommand{\bea}{\begin{eqnarray}}
\newcommand{\eea}{\end{eqnarray}}
\newcommand{\bi}{\bibitem}
\newcommand{\ra}{\rightarrow}

\begin{flushright}
UM - TH - 00 - 13
%\today\\
%hep-ph/0007041
\end{flushright}

\begin{center}
\vglue .06in
{\Large \bf {A Variational Principle for Radial Flows in Holographic Theories.}}\\[.5in]

{\bf R. Akhoury}\\ 
[.05in]

{\it{The Randall Laboratory of Physics\\
University of Michigan\\
Ann Arbor, MI 48109-1120}}\\[.15in]

\end{center}
\begin{abstract}
\begin{quotation}
We  develop furthur the correspondence between a $d+1$ dimensional theory and a $d$
dimensional one with the "radial" $(d+1)$th coordinate $\rho$ playing the role of an 
evolution parameter. We discuss the evolution of an effective action defined on a $d$ dimensional 
surface charactarized by $\rho$ by means of a new variational principle. The conditions under which 
the flow equations are  valid are discussed in detail as is the choice of boundary conditions. 
It is explained how domain  walls may be incorporated in the framework and some generalized 
junction relations are obtained. The general principles are illustrated on the example of a 
supergravity theory on $AdS_{d+1}$.
\end{quotation}
\end{abstract}
\newpage

\section{Introduction }
The holographic principle \cite{hooft,susskind,bousso,suskindreview}, motivated by 
Black Hole considerations, is an extension of the Bekenstein bound \cite{bekenstein}
which limits the number of degrees of freedom that generate entropy in a theory
including gravitation. It is a dynamical statement postulating that the evolution of gravitational
and matter fields in bulk space-time is specified by the data stored in its boundary. In the
conventional quantum field theoretic treatment of gravity, this property is not apparent,
however, recent conjectures in string theory concerning the AdS/CFT duality 
\cite{maldacena, prescription, witten,review}  provide an
example of this principle. If future investigations validate it, the holographic principle could
well turn out to be one of the most important physical ideas in recent times.

There are many forms that a correspondence between a $d+1$ dimensional and a $d$ dimensional 
theory can take. Perhaps the simplest one is that between a classical theory in
the higher dimensional space and a quantum theory at the boundary. An example is provided by 
the conjectured duality between the classical supergravity theory in 5 dimensions and a quantum  
gauge theory in 4 dimensions \cite{maldacena}. More specifically, let $S(\phi,g)$ denote the 
classical action for a supergravity  solution where the boundary values of the fields are $\phi$ 
and of the metric, $g_{\mu\nu}$, then the presciption \cite{prescription,witten} for the conjectured duality is:
\begin{equation}
{1 \over \sqrt{g}}{\delta \over \delta \phi^{i_1}} ...... {1 \over \sqrt{g}}{\delta \over \delta \phi^{i_n}}S
~=~<O_{i_1} ...... O_{i_n}>
\end{equation}
where, $O$ denote certain gauge invariant operators in the boundary theory.  In order for the classical 
description in the bulk of the $AdS_5$ space to  be valid one considers the limit of large N and large 
't Hooft coupling on the gauge theory side. A key ingredient of this correspondence is the UV/IR connection
\cite{uvir1,uvir2} : the infrared regulator of the bulk theory is equivalent to the ultraviolet regulator of the boundary
theory. This indicates that one may interpret the $d+1$th coordinate, $\rho$ as a renormalization group parameter 
of the 4 dimensional theory. Indeed, the evolution of bulk fields with $\rho$ has been studied in a fundamental          
paper \cite{deboer} within the Hamilton-Jacobi framework  where it was shown that 
the resulting flow equations can asymptotically be cast in the form of a Callan-Symanzik equation,
with the bulk scalar field representing the 4 dimensional gauge coupling. In ref. \cite{uvir1} the
boundary AdS/CFT conjectured correspondence has been extended to a finite radius AdS foliation.  
This is of relevance to the  connection with the Randall-Sundrum type scenarios \cite{sundrum}. 
A more generalized correspondence was  originally proposed \cite{maldacena}, i.e., between the type 
II B string theory in the bulk (of $AdS_5$  times $S_5$) and a CFT on the boundary. In the limit of large N 
and large 't Hooft coupling, the string coupling  $g \rightarrow 0$ and one may limit to the classical supergravity
description, otherwise string loops  need to be considered. 

Though much work has gone into the development of these ideas, there are many challenges ahead \cite{review}.
Of particular relevance for this paper is the work of \cite{deboer} concerning the evolution of bulk
fields with $\rho$. We will formulate this evolution 
quite generally by means of a variational principle. The advantages
of doing so are many : a variational principle replaces many mathematical expressions by a single
general principle and moreover the formulation is coordinate independant. The variational principle
which we derive in section 2 involves a multi-stage optimization procedure -a minimization 
condition which leads to a flow equation for the effective action $S$. The effective action is a function of 
the fields on a $d$ dimensional surface charactarized by a particular value of $\rho$. As one changes
$\rho$ the action changes and if we still require a minimum of the action integral, then the flow equations 
must be satisfied. This procedure is discussed in detail in section 2 where its connection with the 
Hamilton-Jacobi formalism is also shown.  The problem is first discussed in a general  setting and 
then the generalized flow equations are  applied to the example of the supergravity theory on 
$AdS_{d+1}$ where the flow equations of ref.\cite{deboer} are reproduced. In addition to the above, a variational 
formulation  provides us with a new way to formulate the boundary  conditions which the flow equations must satisfy.
This is done in section 3 together with a discussion of the Weyl anomaly. Another decided advantage of the variational
formulation is discussed in section 4. There we  discuss in a quantitative manner, the conditions under which the 
"classical" approximation for the  flow equations is valid. In particular, we seek the conditions under which terms
involving the second partial derivatives of the effective action which are ignored in the derivation of the flow equations
in section 2 can  become unbounded. This allows us to formulate the applicability criterion quite generally
in terms of (the solutions of) a set
of partial differential equations. The example of a single scalar field in AdS is then considered where the
differential equation reduces to a Jacobi type equation. The question of how the second partial derivatives 
of the action may be incorporated into the flow equations in this framework will be dealt with in a subsequent
publication \cite{akh}. Our formulation, as with ref.\cite{deboer}, allows one to consider
the effective action for values of $\rho$ away from the boundary. In the AdS case, for example, we can consider
any of its foliations. This allows us to study the scenario when a domain wall is introduced into the
$d+1$ dimensional space-time. This is done quite generally in section 5 where starting from the results 
of section 2, generalized junction conditions \cite{israel} are derived. The example of the Randall-Sundrum
\cite{sundrum}  scenario and its connection with the supergravity theory on AdS \cite{connection}
is also discussed in some detail. Finally, in section 6 we conclude with a discussion of the results. 

\section{The Variational Principle}
In this section we will consider the dynamical evolution of fields in a $d+1$ dimensional spacetime
with a $d$ dimensional (time-like) boundary.  We will denote the $d$ dimensional coordinates by
$x_{\mu}$ and the "radial" coordinate is $\rho$.  We consider a $d$ dimensional hypersurface defined by
$\rho =$ constant after making appropriate gauge choices for the metric. In the AdS case, for example,
this hypersurface would be a particular foliation. The boundary of the
$d+1$ dimensional space is located at $\rho =\rho_0$. We will be interested in fields $\phi_i(x,\rho)$
\cite{note1} whose dynamics is governed by a lagrangian $\cal{L}$. In particular, we would like
to understand how the effective action changes as we move from one hypersurface to a neighbouring
one. The trajectory that one follows in moving from one hypersurface to the other is not arbitrary
but is determined by an optimization condition.   We propose below that this optimal trajectory is determined 
by a variational principle which will be seen to imply a (functional) partial differential equation satisfied by 
the effective action $S({\phi_i})$ of  the boundary values  of the fields. 

The optimal trajectory arises as a result of a minimization problem and the solution suggests itself 
as a consequence of a multi-stage procedure. To see how this works, consider  first the integral:
\be
J'[\phi_i]=\int_{\rho_1}^{\rho_2}d\rho F(\rho, \phi_i,\dot{\phi_i})
\ee
In the above, $\dot{\phi_i}=d\phi_i/d\rho$. In general, $F(\rho, \phi_i,\dot{\phi_i})$ will be of
the form, $\int_d\sqrt{G}$\( {\cal L} (\rho, \phi_i,\dot{\phi}_i) \), with $\cal{L}$ the Lagrangian, 
and $G_{\mu\nu}$ is the $d+1$ dimensional metric. In this and the following we
suppress  any dependance on the $d$ dimensional coordinates. We will call the critical trajectory,
the one which is obtained by minimizing $J'$. Consider a trajectory, and let 
us choose a point $(\rho_3,\phi_{i3})$ \cite{note2} in between the initial and final points, 
and follow the curve from $(\rho_1,\phi_{i1})$ to
$(\rho_3,\phi_{i3})$. For the rest of the curve ($(\rho_3,\phi_{i3})$ to $(\rho_2,\phi_{i2})$ to be also critical we
must minimize $\int_{\rho_3}^{\rho_2}d\rho F(\rho, \phi_i,\dot{\phi_i})$. This must be true for all such
inbetween points $\rho_3$. Thus whatever the initial point $(\rho_n,\phi_{in})$, or the initial arc of the trajectory, 
the remaining transformations must constitute an optimal sequence for the remaining problem.  The endpoint
of the first transformation is the initial point of the next and so on. The key point here is that the perturbation
in the first interval produces a dependant deformation of the remaining curve. Let us next apply such a
multi-stage procedure to the variational problem of minimizing $J'$.

In view of the above comments, let us consider instead of Eq. (1), the following integral 
\be
J[\phi_i]=\int_{\rho_1}^{\rho}d\rho' F(\rho', \phi_i(\rho'),\dot{\phi_i}(\rho')).
\label{j}
\ee
Suppose that for a given $(\rho,\phi_i)$ there is some trajectory that minimizes $J$. Let us denote
\be
S(\rho,\phi_i)=Min \left\{ \int_{\rho_1}^{\rho}d \rho' F(\rho', \phi_i(\rho'),\dot{\phi_i}(\rho')) \right\}
\label{defS}
\ee
In order to use the idea presented above, we first divide the interval thus:
\be
(\rho_1,\rho) = (\rho_1,\rho-\Delta \rho) + (\rho-\Delta \rho,\rho).
\ee
Along the arc associated with the first interval we take the trajectory to be critical and for the second
interval it is, in general, arbitrary except that at the endpoint $\phi_i(\rho)=\phi_i$. The corresponding 
contributions to the integral $J$ are denoted by $J_1$ and $J_2$;
\be
J_1= \int_{\rho_1}^{\rho-\Delta \rho}d\rho' F(\rho', \phi_i(\rho'),\dot{\phi}_i(\rho'))
\ee
and,
\be
J_2= \int_{\rho-\Delta \rho}^{\rho}d\rho' F(\rho', \phi_i(\rho'),\dot{\phi}_i(\rho')).
\ee
Since the trajectory is critical in the first interval, we get using the definition in Eq.(\ref{defS}),
\be
J_1 = S(\rho - \Delta \rho, \phi_i(\rho - \Delta \rho))
\ee
Expanding to first order in $\Delta \rho$,
\be
J_1 = S(\rho-\Delta \rho, \phi_i - \dot{\phi}_i\Delta \rho)
\label{defj1}
\ee
An important point to notice here is that the function $\dot{\phi_i}$ is arbitrary above and in the following
expression for $J_2$. Since the second interval is infinitesmal, we have again to first order in $\Delta \rho$
\be
J_2 = F(\rho, \phi_i,\dot{\phi_i}) \Delta \rho.
\label{defj2}
\ee
Since the sum of $J_1$ and $J_2$ is greater than or equal to $S$, and since in Eq.(\ref{defj1}) 
and Eq.(\ref{defj2}) $\dot{\phi}_i$ is the arbitrary function, we have:
\be
S(\rho,\phi_i) = \stackrel{\textstyle Min}{\dot{\phi}_i} \left\{ F(\rho, \phi_i,\dot{\phi}_i) \Delta \rho +
 S(\rho-\Delta \rho, \phi_i - \dot{\phi}_i\Delta \rho) \right\}
\ee
Assuming that the second and higher partial derivatives of S are bounded we can again neglect 
higher orders in $ \Delta\rho$ to get, 
\be
S(\rho,\phi_i) = \stackrel{\textstyle Min}{\dot{\phi}_i}\left\{ F(\rho, \phi_i,\dot{\phi_i}) \Delta \rho + 
S(\rho,\phi_i) - {\partial S \over \partial \rho}\Delta \rho -
{\delta S \over \delta \phi_i}\dot{\phi_i} \Delta \rho \right\}.
\ee
Finally, in the limit of $\Delta \rho \rightarrow 0$ we obtain:
\begin{equation}
\stackrel{\textstyle Min}{\dot{\phi}_i}\left\{ F(\rho, \phi_i,\dot{\phi_i}) - {\partial S \over
\partial \rho}-{\delta S \over \delta \phi_i}\dot{\phi}_i \right\} = 0.
\label{master}
\end{equation}
This is the main result of this section.  It is worth emphasizing here that, $\dot{\phi_i}(\rho,\phi_i)$, the value
of which at each point $(\rho,\phi_i)$ minimizes the expression on the right hand side of the above equation is
associated with a solution $S(\rho,\phi_i)$ of that equation.  If $S$ is known then its partial derivatives
are easily obtained  and the value of the "flow velocities" $\dot{\phi_i}(\rho,\phi_i)$ at any point can be
determined by minimizing a function of only the variables $\dot{\phi_i}$; conversely, if
$\dot{\phi_i}(\rho,\phi_i)$ are known everywhere then $S$ can be obtained by evaluating the integral $J$ in
Eq.(\ref{j}) with the given flows. Thus knowledge of the flow velocities or the action along the optimal
trajectory constitutes a complete solution to the problem.

From the above,
it follows that at the minimum,
\begin{equation}
F(\rho, \phi_i,\dot{\phi_i}) = {\partial S \over \partial \rho}+{\delta S \over \delta \phi_i}\dot{\phi_i}
\label{hj}
\end{equation}
and in addition,
\begin{equation}
{\delta F \over \delta \dot{\phi_i}} = {\delta S \over \delta \phi_i}
\label{hj2}
\end{equation}
It should be noted that Eq.(\ref{hj2}) asserts that for a given numerical value of $\rho,\phi_i, {\partial S \over
\partial \rho},{\delta S \over \delta \phi_i}$, the derivative w.r.t. $\dot{\phi_i}$ of Eq.(\ref{master})
must be zero when $\dot{\phi_i}$ minimizes the bracketed quantity. The above two equations imply the
Hamilton Jacobi equation $(H+{\partial S \over \partial \rho}=0)$, for this system as can be easily 
seen by noting that the Hamiltonian is given by:
\begin{equation}
H = -F + \dot{\phi_i}{\delta F \over \delta \dot{\phi_i}}.
\end{equation}
In theories including gravity,  since the local shifts $\rho \rightarrow \rho+\Delta \rho$ are part of the
general coordinate invariance, we have the implied constraint , $H = 0$. In this case it is the following 
(functional) differential equation which determines the action functional $S$:
\begin{equation}
F(\rho, \phi_i,\dot{\phi_i}) = {\delta S \over \delta \phi_i}\dot{\phi_i}
\label{hj1}
\end{equation}
and in addition the momenta conjugate to the fields are given by:
\begin{equation}
\Pi_i = {1 \over \sqrt{g}}{\delta F \over \delta \dot{\phi_i}}
\label{mom}
\end{equation}
Eqs.(\ref{hj2},\ref{mom}) allow us to relate the momenta with the flow velocities $\dot{\phi_i}$ of the
fields.  Since the effective action, $S=S(\phi_i,\rho)$ only, we see that these equations also imply that we may in
principle solve for the $\dot{\phi}_i$ in the form:
\begin{equation}
\dot{\phi}_i~=~g_i(\phi_i, {\delta S \over \delta \phi_i}, \rho)~=~f_i(\phi_i,\rho).
\label{dotphi}
\end{equation}

Next we briefly discuss the relation of the fundamental Eq.(\ref{master}) or Eqs.(\ref{hj}, \ref{hj2}) to the
$d+1$-dimensional equations of motion.  Henceforth we will adopt the notation (the derivatives may be functional 
or ordinary partials), 
\begin{equation}
S_{,y_1y_2}~\equiv~{\delta^2 S \over \delta y_1 \delta y_2},
\end{equation}
and so on. Along the critical trajectory, consider
\begin{equation}
\left(d S_{,\phi_i}/d\rho \right)_{\dot{\phi}_k}~=~S_{,\rho \phi_i} ~+~ S_{,\phi_i \phi_j}\dot{\phi}_j
\label{euler}
\end{equation}
Similarly, differentiating Eq.(\ref{hj}) with respect to $\phi_i$ gives,
\begin{equation}
F_{,\phi_i} + (F_{,\dot{\phi}_j} - S_{,\phi_j}){\delta \dot{\phi}_j \over \delta \phi_i} - S_{,\rho \phi_i}
- S_{,\phi_i \phi_j}\dot{\phi}_j ~=~0.
\label{euler2}
\end{equation}
We will discuss the possibility of $\dot{\phi}_i$ being discontinuous along a critical trajectory in a later
section, hence omitting this possibility for the moment and using Eq.(\ref{hj2}) and Eq.(\ref{euler}) we get,
\begin{equation}
(d S_{,\phi_i}/d\rho)_{\dot{\phi_k}}~=~F_{,\phi_i}.
\label{euler2}
\end{equation}
In a similar manner and under similar assumptions we can also get:
\begin{equation}
(d S_{,\rho}/d\rho)_{\dot{\phi_k}}~=~F_{,\rho}
\label{euler3}
\end{equation}
Using Eq.(\ref{hj2}), it is seen that Eq.(\ref{euler2}) is in fact the $d+1$ dimensional Euler Lagrange equation:
\begin{equation}
(d F_{,\dot{\phi}_i}/d\rho)_{\dot{\phi_k}}~=~F_{,\phi_i},
\end{equation}
which are seen to be satisfied along the critical trajectory.

We have seen that corresponding to the critical trajectory there exist partial derivatives of
the action $S$ which satisfy the differential equations (\ref{euler2},\ref{euler3}) such that the flow velocity
along the critical trajectory minimizes for each value of $\rho$ the expression :
\begin{equation}
F~-~S_{,\rho}~-~S_{,\phi_i}\dot{\phi}_i,
\end{equation}
and in addition that minimum value is zero. This allows for an interpretation which is in the spirit of the
renormalization group approach. The "cut-off" corresponds to a particular value of $\rho$. As we change 
the cut-off, we continue to require that  the trajectory is critical and this gives us
Eq.(\ref{master}). The variational principle thus can be thought of as an alternative formulation of the
renormalization group flow. However up until now we are restricting ourselves to the classical theory in the bulk.
As discussed in Sec.[4]  the above treatment breaks down when the second derivatives of $S$ cannot be
neglected in the derivation leading upto Eq.(\ref{master}). Then a more general (quantum) treatment is
required, which is developed in  ref.\cite{akh}. As mentioned earlier, in $d+1$ dimensional theories with
gravity we must set the gauge constraint: $S_{,\rho}~=~0$. In such theories, of interest to the holographic
principle, the relevant equations are (\ref{hj1}), (\ref{hj2}) and (\ref{mom}). 

To summarize let us express the results of this section in a geometrical setting. At all points other than the 
initial point the hypersurface (foliation) in the $d+1$ dimensional space may be denoted by:
\begin{equation}
S(\phi_i, \rho)~=~\alpha,
\end{equation}
where $\alpha$ is a parameter.
We have also seen that along the critical trajectory, Eq.(\ref{hj2}), is satisfied. As discussed earlier, 
it allows us to solve for $\dot{\phi}_i$ through Eq.(\ref{dotphi}). The curves so defined intersect the 
hypersurface, the former being determined by the latter. Let $P(\phi_i,\rho)$ denote a point on the 
hypersurface $S=\alpha$. A critical curve passes through $P$ and will intersect the neighbouring 
hypersurface $S=\alpha + d\alpha$ at some point $Q(\phi_i + \delta \phi_i, \rho + \delta \rho)$, where
$\delta \phi_i=\dot{\phi}_i d\rho=f_i(\phi_i, \rho) d\rho$. The displacement $PQ$ is, $d\alpha=F d\rho$,
which is independant of the position of $P$ on the first hypersurface $S=\alpha$. Thus along the critical
trajectory when we go from one hypersurface to another, the increment of the fundamental integral 
is always $d\alpha$. Generalizing this we may integrate from $S=\alpha_1$ to $S=\alpha_2$ along the
critical trajectory :
\begin{equation}
\int_{P_1}^{P_2}F d\rho~=~\int_{P_1}^{P_2} \left( {\partial S \over \partial \rho} +
 {\delta S \over \delta \phi_i}\dot{\phi}_i \right)~=~\alpha_2 - \alpha_1,
\end{equation}
which is independant of the position of $P_1$. Thus we have an interesting result that 
a family of critical curves will cut off "equal distances" between two such surfaces.

In order to illustrate the above ideas let us consider in some detail a specific example in the framework of
the $AdS_{d+1}/CFT_{d}$ correspondence. As usual, we will consider a supergravity theory on $AdS_{d+1}$ 
which contains scalars $\phi^I$. This supergravity theory is the low energy limit of the type $IIB$ string 
theory  and the boundary CFT it is conjecturaly equivalent to is a ${\cal{N}}= 4$ SYM theory in the limit of 
large $N$ and large 't Hooft coupling. The action for the bulk theory is
(setting the $d+1$ dimensional Newton's constant to unity) :
\begin{equation}
I_{d+1}~=~\int_{d+1} (R^{d+1} + 2\Lambda) + 2\int_d K + \int_{d+1} (V(\phi) + \frac{1}{2} G_{IJ} \nabla_a 
\phi^{I} \nabla^a \phi^{J}),
\label{adslag}
\end{equation}
where, the $d+1$ dimensional cosmological constant $2\Lambda \equiv {d(d-1) \over r^2}$, and $r$
is the AdS radius, $V$ and $G_{IJ}$ are
the $d+1$ dimensional scalar potential and metric and $K$ is the trace of the extrinsic curvature, $K_{\mu\nu}$
of an arbitrarily chosen foliation at $\rho =$ constant. For reasons of uniqueness of solutions \cite{witten}
we henceforth take the $d+1$ dimensional space to have Euclidean signature with a metric :
\begin{equation}
ds^2~=~d\rho^2 + g_{\mu\nu}(x,\rho)dx^{\mu}dx^{\nu},
\label{gauge}
\end{equation}
which entails a specific gauge choice. For large $\rho$ the boundary metric $g_{\mu\nu} \sim e^{2\rho/r}$ so
infinities arise as we take the boundary to infinity. these must be cancelled with counterterms 
 \cite{witten,hensken,tseytlin, myers}
before the limit is taken. It has been dicussed previously that crucial to the duality of the bulk and boundary thoeries is
the UV/IR connection. With the choice of gauge (\ref{gauge}), the extrinsic curvature of a foliation at $\rho$ is, 
\begin{equation}
K_{\mu\nu}~=~\frac{1}{2}\dot{g}_{\mu\nu}.
\end{equation}
Using the Gauss-Cordacci equations one obtains :
\begin{equation}
I_{d+1}~=~\int d^dx d\rho \sqrt{g}(R^{d} + K^2 - K_{\mu\nu}K^{\mu\nu} + 2\Lambda + V(\phi) 
+ \frac{1}{2} G_{IJ}\nabla_a \phi^{I} \nabla^a \phi^{J}).
\end{equation}
With the identification,
\begin{eqnarray}
F~=~\int d^dx(\sqrt{g}\cal{L}) \\
\sqrt{g}{\cal{L}}~=~\sqrt{g}(R^{d} + K^2 - K_{\mu\nu}K^{\mu\nu} + 2\Lambda + V(\phi) 
+ \frac{1}{2} G_{IJ}\nabla_{\mu} \phi^{I} \nabla^{\mu} \phi^{J});
\end{eqnarray}
the canonical momenta conjugate to $g^{\mu\nu}$ and $\phi^I$ are,
\begin{eqnarray}
\pi_{\mu\nu}~=~\frac{1}{\sqrt{g}}{\delta F \over \delta \dot{g}^{\mu\nu}}~=~K_{\mu\nu} - K g_{\mu\nu}\\
\pi_I~=~\frac{1}{\sqrt{g}}{\delta F \over \delta \dot{\phi}^{I}}~=~G_{IJ}\dot{\phi}^J .
\end{eqnarray}
The above and equations (\ref{hj2}) now give the flow velocities as :
\begin{eqnarray}
\dot{g}_{\mu\nu}~=~\frac{2}{\sqrt{g}}{\delta S \over \delta {g}^{\mu\nu}} - \frac{2}{d-1}g_{\mu\nu}
 g^{\alpha\beta}\frac{1}{\sqrt{g}}{\delta S \over \delta {g}^{\alpha\beta}}  \\
\dot{\phi}^I~=~G^{IJ}\frac{1}{\sqrt{g}}{\delta S \over \delta \phi^{J}}. 
\end{eqnarray}
The master equation (\ref{hj1}) for this case can hence can be written as,
\begin{equation}
\frac{1}{\sqrt{g}}\left(-{\delta S \over \delta g_{\mu\nu}}{\delta S \over \delta g^{\mu\nu}} -
\frac{1}{2}{\delta S \over \delta \phi^I}{\delta S \over \delta \phi^J}G^{IJ} +
\frac{1}{d-1}\left(g^{\mu\nu}{\delta S \over \delta g^{\mu\nu}}\right)^2 \right)~=~\sqrt{g} (R^{d} +
2\Lambda + V(\phi) +\frac{1}{2} G_{IJ}\nabla_{\mu}\phi^{I}\nabla^{\mu} \phi^{J})
\label{adsmaster}
\end{equation}
This equation, first obtained in \cite{deboer}, determines the form of the effective action as a function 
of the boundary fields $\phi_I$ and
$g_{\mu\nu}$ when the boundary is at $\rho$. Changing $\rho$ changes the boundary and the effective action also
changes in such a manner that one is still on a critical trajectory. As emphasized before, the $d+1$ coordinate
$\rho$ plays the role of a renormalization group parameter and it was shown \cite{deboer} that in the asymptotic 
limit the above equation can be cast in the form of a Callan-Symanzik equation. This provides justification for the
identification of $S$ with the quantum effective action of a $d$ dimensional theory at the boundary of
$AdS_{d+1}$.

Equation(\ref{adsmaster}) can be used to determine the form of the effective action. In general, at some energy
scale, $S$ may be split into  a local and a non-local part and a derivative expansion of the former performed . The
different terms in the derivative expansion are constrained by (\ref{adsmaster}), and in fact thier coefficients 
can be thus determined. Including terms upto four derivatives we can write an ansatz for $S_{loc}$ as :
\begin{equation}
S_{loc}~=~\int_{d}\sqrt{g}\left(\Phi_1(\phi)R + U(\phi) + \Phi_2(\phi)R^2 + \Phi_3(\phi)R_{\mu\nu}R^{\mu\nu} +
\Phi_4(\phi)R_{\mu\nu\alpha\beta}R^{\mu\nu\alpha\beta} \right)
\label{sloc}
\end{equation}
Then we get for the coefficients of the leading two terms in (\ref{sloc}),
\begin{eqnarray}
U^2 - {4(d-1) \over d}\frac{1}{2}G^{IJ}U_{,IJ}~=~{4(d-1) \over d}2 \Lambda  + {4(d-1) \over d}V  \label{U} \\
{d-2 \over 2(d-1)} U\Phi_1 - G^{IJ}U_{,IJ}~=~1
\label{phi} 
\end{eqnarray}
In addition, for the case of pure gravity ($V=0$, $\delta_{\mu} \phi_I=0$) we get the following for the 
coefficients of the higher derivative terms \cite{note4},
\begin{eqnarray}
\Phi_2~=~-\frac{1}{2}{d \over d-4}{\Phi_1^2 \over U}, \\
\Phi_3~=~2{d-1 \over d-4}{\Phi_1^2 \over U}, \\
\Phi_4~=~0.
\end{eqnarray}
We will make use of these results in a future section. 

\section{Boundary Conditions.} 
In the context of holographic theories, specially the AdS/CFT example, several types of boundary conditions  have
been proposed for the fields on the boundary. In this section we will consider this question from the perspective
of the differential equations (\ref{hj}, \ref{hj1}) and (\ref{hj2}), for the effective action $S$. The proper choice 
of the boundary conditions on $S(\rho, \phi_i)$, in contrast to Eqs.(\ref{hj}, \ref{hj1}), does depend upon the 
initial conditions  on the allowed trajectories.   We will consider some of the possibilities below.

If the allowed trajectories are those that start at any point on the "surface" (read foliation for the AdS
example) $\rho=\rho_1$, then if the function
$F(\rho, \phi_i,\dot{\phi_i})$ does not develop a divergence at that point, we have
$S(\rho_1, \phi_i) = 0$ for all $\phi_i$. This follows from Eq.(\ref{defS}) where at $\rho=\rho_1$, the upper and
lower limits  coincide.  Note that for this choice of boundary conditions $\phi_i(\rho_1)$ is unspecified. It is
also important to realize that with this choice of boundary conditions, $S_{,\phi_i}$ and
$S_{,\phi_i \phi_j}$ are also identically zero along the "surface" $\rho = \rho_1$. We should mention
a straightforward generalization of this case, i.e., if the
initial condition is such that $\phi_i(\rho_1)=f_i(\rho_1)$, then $S(\rho,\phi_i)=0$ for all $\rho,\phi_i$ such
that $\phi_i=f_i(\rho)$, for the same reasons as before. 

We next discuss a different choice of boundary conditions which is more relevant to  applications of the
holographic principle. This is the situation that all allowed critical curves begin at a specified initial point
($\rho_1, \phi_i(\rho_1)$). Since $\phi_i(\rho_1)$ is specified, no other points
at $\rho =\rho_1$ with $\phi_i$ having any other value is allowed. Then, as $S$ is defined along
the "surface" $\rho =\rho_1$ only at $\phi_i = \phi_{i1}$ at which point $S(\rho_1, \phi_{i1}) =
0$, we see that the effective action does not have a well defined partial derivative with respect
to $\phi_i$ at this point. Along the critical curve, in fact, $S_{,\phi_i}$ is well defined  at all
$(\rho, \phi_i)$ except at the initial point.  As we approach an arbitrarily close final point from
the initial point, along the critical curve, $S_{,\phi_i}$ is well defined and as will be shortly
seen, this allows us to obtain a limiting value for this quantity at
$(\rho_1, \phi_{i1})$. Of course, this limiting value of $S_{,\phi_i}$ will be different for different critical
curves emanating from the same initial point but ending at different final points. 

Indeed, we have seen in the previous section that along the critical curve, Eq.(\ref{hj2}) is
satisfied. This is true at all points except at the initial point where $S_{,\phi_i}$ is not well
defined. However, we can extrapolate back from an arbitrarily close point along the curve to the
initial point assuming a "straight -line" path. Thus from a point $(\rho, \phi_i)$ very close to the
initial point, $(\rho_1, \phi_{i1})$ we approximate the critical curve by a "straight line" and
write:
\begin{equation}
\dot{\phi}_i~=~{\phi_i - \phi_{i1} \over \rho - \rho_1}.
\label{slope}
\end{equation}
Now we can evaluate $F_{,\dot{\phi}_i}$ using the above, and demand that Eq.(\ref{hj2}) is
satisfied as the reference point $(\rho, \phi_i)$ approaches the initial point. The picture of
the allowed trajectories that thus emerges is the following: The critical curves all start from
the same initial point $(\rho_1, \phi_{i1})$ and fan out each with different values of the
initial "slope" $S_{,\phi_i}$ arriving at a boundary manifold at different points. As seen from the
boundary  the various critical trajectories come from a "focal point" (the initial point) in the bulk
of the $d+1$ dimensional space-time. 

The limiting behaviour of ${\delta \dot{\phi}_i \over \delta \phi_j}$ as $(\rho, \phi_i)$
approaches the initial point can also be obtained for this choice of boundary conditions from the
following argument : Expanding $\phi_{i1}$ in a Taylor series, about $\phi_i(\rho)$ which is
close to it, we have,
\begin{equation}
\phi_i(\rho_1)~=~\phi_i(\rho) + (\rho_1 - \rho)\dot{\phi}_i(\rho) + ........
\end{equation}
We now replace functional derivatives with ordinary derivatives by using the definition,
\begin{equation}
\int{\delta \over \delta \phi_i}~=~{\partial \over \partial \phi_i}.
\label{partial}
\end{equation}
Since $\phi_{i1}$ is fixed, we get upon differentiating the above,
\begin{equation}
0~=~\delta_{ij} + {\partial \dot{\phi}_i \over \partial \phi_j}(\rho_1 - \rho) + .....
\end{equation}
This implies that,
\begin{equation}
{\partial \dot{\phi}_i \over \partial \phi_j} ~ \sim ~{\delta_{ij} \over \rho -\rho_1} + {\cal{O}} (1).
\end{equation}
This singularity at the initial point is not surprising and is just a reflection of the near "straight
-line" behaviour for points very close to the initial point. The problems with the unboundedness of
${\partial \dot{\phi}_i \over \partial \phi_j}$ arise if it has a singularity at any other point
than the initial one. This is discussed in the next section.

Which boundary condition one adopts, depends on the problem under consideration.
In holographic theories, as discussed earlier, the correspondence between the bulk and boundary
theories is  given by, 
 \begin{equation}
< O_i.....O_j >~ \sim~ {\delta \over \delta\phi_i}.....{\delta \over \delta\phi_j}S(\phi)
\end{equation}
where, $\phi_i$ are the boundary values of the fields and $O$ are the invariant operators in the 
 boundary theory. Thus, the proper choice of boundary conditions in such theories should be
such that these derivatives of $S$ are nonvanishing. In general, specifying the conjugate
momenta of the fields in the bulk amounts to specifying the nature of the vaccum. For
gravitational theories the conjugate momenta relative to $g_{\mu\nu}$ is the energy momentum
tensor. We have mentioned earlier that different possible trajectories emnate from the initial
point with the slopes ${\delta S \over \delta \phi_i}$ being the distinguishing charactaristic.
From a physical point of view, picking a particular value for the slope amounts to choosing a
particular vaccum and expectation value of the energy momentum tensor. In fact, in theories
including gravitation, it is only the last choice of boundary conditions which ensures even the
possibility of a nonvanishing Weyl anomaly. Indeed, it is easy to relate the rate of change of the
effective action $S$ (see eq.(\ref{defS})) to the Weyl anomaly; 
\begin{equation}
{d S \over d \rho}~=~{\partial S \over \partial \rho} + {\delta S \over \delta \phi_i}\dot{\phi}_i.
\label{schange}
\end{equation}
This is just Eq.(\ref{master}) written in another form. In theories with gravitation, the first
term on RHS vanishes and
\begin{equation}
{d S \over d \rho}~=~{\delta S \over \delta g_{\mu\nu}}\dot{g}_{\mu\nu}.
\label{weylanomaly}
\end{equation}
where for simplicity we have only included the gravitational field. Using (\ref{mom}) we
may in principle solve for the flow velocity, $\dot{g}_{\mu\nu}$ in terms of the metric and substitute in 
the above equation. In practice, however this solution gives a non-local expression on the RHS of
(\ref{weylanomaly}). In some local approximation, the solution of the flow velocity in terms of
the metric may be used to obtain the Weyl anomaly at the boundary since then (see example below),
\begin{equation}
RHS (\ref{weylanomaly})~\sim~\int_d <T_{\mu}^{\mu}>,
\end{equation}
with $<T_{\mu}^{\mu}>$ the trace of the energy momentum tensor. We note that for practical calculations it
is easy to evaluate the LHS of (\ref{weylanomaly}) in some local approximation.
Let us briefly indicate how this works \cite{note3} for the conjectured AdS/CFT duality from the bulk point
of view.

For the purpose of studying the Weyl anomaly using this method, for the AdS example it will be
convenient (though by no means essential) to make a different gauge choice from the one given in (\ref{gauge}). 
We will restore the lapse function $N$ and choose a gauge with,
\begin{eqnarray}
ds^2~=~N^2 d\rho^2 + g_{\mu\nu}dx^{\mu}dx^{\nu} \\
N~=~{r \over 2\rho}
\label{gauge2}
\end{eqnarray}
In this gauge the AdS boundary is at $\rho = 0$ and the appropriate changes in the equations 
at the end of the previous section are easy to trace. The most relevant change for us is
$\dot{g}_{\mu\nu} \rightarrow N^{-1} \dot{g}_{\mu\nu}$ and this implies that in the case
without the scalar fields, the flow and the master equations give in particular,
\begin{equation}
U^2~=~{4 \over N^2}{d-1 \over d}2\Lambda.
\end{equation}
Then, in the approximation when the scalar potential $U$ dominates in $S$, 
\begin{equation}
\dot{g}_{\mu\nu}~=~{2N \over r} g_{\mu\nu}~=~{1 \over \rho} g_{\mu\nu}.
\end{equation}
Eq.(\ref{weylanomaly}) with $\phi_i = g_{\mu\nu}$ in the above mentioned approximation can
now can be written as,
\begin{equation}
\rho{d S \over d \rho}~=~g_{\mu\nu}{\delta S \over \delta g_{\mu\nu}}~=~\int_d <T_{\mu}^{\mu}>,
\end{equation}
 It is east to check from the results of
ref.\cite{hensken} that the LHS of the above equation, evaluated at the boundary ($\rho \rightarrow 0$), 
correctly reproduces the Weyl anomaly there. Note that in general some local invariant counterterms 
must be added to remove infinities in the action.

It is interesting that a classical calculation provides us with an answer which from the point of view of the
boundary theory is purely quantum mechanical. We may attempt to understand this drawing from Dirac's
connection between the classical Poisson brackets (PB) and the quantum mechanical commutator, i.e.,
\begin{equation}
[ u, v ]~=~i\frac{h}{2\pi}\{ u, v \}_{PB}
\end{equation}
 Consider Eq.(\ref{schange}) for the gravity case (${\delta S \over \delta \rho}=0$), and substitute for 
$\dot{\phi}_i = {\delta H \over \delta \Pi^i}$. Then noting that $S$ only depends on $(\rho,\phi_i)$
this equation becomes :
\begin{equation}
{d S \over d \rho}~=~\{ S, H \}_{PB}.
\end{equation}
Thus, what we identify with the Weyl anomaly at the boundary in the local limit is :
\begin{equation}
 \lim_{\rho \rightarrow 0} \rho \{ S, H \}_{PB} .
\end{equation}
The strong Poisson bracket and commutator analogy gives us the required result. The question of quantum 
corrections and the connection with the stochastic quantization method will be discussed in detail in a separate
publication \cite{akh}.

\section{Limits of Applicability.}  
In the previous sections we have obtained  the equations that govern the
effective action and determine its functional form. In deriving the master equation, (\ref{master}), we assumed
the existence and the boundedness of the second partial derivatives of $S$; in fact, in the derivation we
neglected the contributions of the second order derivatives compared to the first. In this section we will
study the conditions when this is true. The unboundedness of the second partial derivatives of $S$ imply that the
"classical" treatment of this paper is not a good enough approximation. The complete bulk quantum mechanical
corrections involving the second derivatives of $S$ would now have to be added on to (\ref{master}). How this can
be implemented is the subject of a separate publication.

It is clear that when the initial point is in the neighbourhood of the terminal point , the master equation
(\ref{master}), is always valid. As the $\rho$ interval increases, it is possible that the solution to
(\ref{master}) may undergo a qualitative change.  It is at such a point that the second derivative of $S$ 
can become
unbounded. In order to quantitatively address this problem, we first relate the unboundedness of the second
partial derivatives of $S$ to that of the derivative of the flow velocities and then obtain a differential equation
for  the latter which does not involve the effective action $S$ but only the "tree" level quantity $F$ and its partial
derivatives. Again, throughout the following discussion of boundedness we will replace functional derivatives 
with ordinary derivatives using Eq.(\ref{partial}) as required.

Indeed, partial differentiation of Eq.(\ref{hj2}) w.r.t. $\phi_i$ gives:
\begin{equation}
F_{,\phi_i \dot{\phi}_j}~+~ F_{,\dot{\phi}_k\dot{\phi}_j}{\partial \dot{\phi}_k \over \partial \phi_i}
~=~S_{,\phi_i \phi_j}.
\label{u1}
\end{equation}
Thus, $S_{,\phi_i \phi_j}$ will be unbounded if (a)~ $F_{,\phi_i \dot{\phi}_j}$~ is unbounded,
(b)~$F_{,\dot{\phi}_k\dot{\phi}_j}$~ is non-zero and ${\partial \dot{\phi}_k \over \partial \phi_i}$ is
unbounded. Case (a) is trivial and we will focus on case (b) assuming that the second partial derivatives of $F$
are bounded. Then at those points where ${\partial \dot{\phi}_k \over \partial \phi_i}$ becomes infinite our
procedure breaks down and Eqns.(\ref{master}, \ref{hj}, \ref{hj2}) are no longer valid. Higher order (quantum)
 corrections
must be included in the form of the second partial derivatives of $S$. We will now attempt to understand
this breakdown more precisely and to discuss the conditions for it to occur.

Having exchanged the problem of unboundedness of the second partial derivative of $S$ for that of the
derivative of the flow velocity, we would like to remove all reference to the action $S$ and its derivatives 
in favor of the known $F$ and its partial derivatives. To this end consider the partial derivative of
Eq.(\ref{hj}) w.r.t. $\phi_i$ for the critical trajectory,
\begin{equation}
F_{,\phi_i}~=~S_{,\rho \phi_i} + S_{,\phi_i \phi_j}\dot{\phi}_j,
\label{u2}
\end{equation}
where we have used Eq.(\ref{hj2}) once to cancel terms involving the derivative of the flow velocities.
Next from Eq.(\ref{u1}) we get by taking its total derivative w.r.t. $\rho$,
\begin{equation}
\left({d \over d\rho} \left( F_{,\phi_i \dot{\phi}_j} + F_{,\dot{\phi}_k\dot{\phi}_j}{\partial \dot{\phi}_k
\over \partial \phi_i} \right) \right)_{\dot{\phi}_l}~=~S_{,\rho \phi_i \phi_j} + S_{,\phi_i \phi_j \phi_k}
\dot{\phi}_k.
\label{u3}
\end{equation}
From Eq.(\ref{u2}) we have :
\begin{equation}
F_{,\phi_i \phi_j} + F_{,\phi_i\dot{\phi}_k}{\partial \dot{\phi}_k \over \partial \phi_j}~=~S_{,\rho
\phi_i \phi_j} + S_{,\phi_i \phi_j \phi_k}\dot{\phi}_k + S_{,\phi_i \phi_k}{\partial \dot{\phi}_k \over  
\partial \phi_j}
\label{u4}
\end{equation}
Eqns.(\ref{u2}, \ref{u3}, \ref{u4}) now give:
\begin{equation}
\left({d \over d\rho} \left( F_{,\phi_i \dot{\phi}_j} + F_{,\dot{\phi}_k\dot{\phi}_j}{\partial \dot{\phi}_k
\over \partial \phi_i} \right) \right)_{\dot{\phi}_l} - F_{,\phi_i \phi_j} + F_{,\dot{\phi}_i \dot{\phi}_l}
{\partial \dot{\phi}_l \over \partial \phi_k}{\partial \dot{\phi}_k \over \partial \phi_j}~=~0.
\label{umaster}
\end{equation}
Eq.(\ref{umaster}) is the principal result of this section. 

To proceed furthur with an analysis of these equations, we need to specify the boundary conditions for the
problem. Suppose that the problem under consideration requires the use of boundary conditions such that 
the critical curves start at a specified point in $(\rho, \phi_i)$ space, i.e., at
$\rho = \rho_1$, $\phi_i = \phi_{i1}$. Then as discussed in the previous section, near the initial point
$\rho \rightarrow \rho_1$, 
\begin{equation}
{\partial \dot{\phi}_i \over \partial \phi_j}~=~-{\delta_{ij} \over \rho_1 -\rho} +{\cal{O}}(1).
\end{equation}
Thus ${\partial \dot{\phi}_i \over \partial \phi_j}$ is singular at the initial point. We have seen that
Eq.(\ref{umaster}) describes the behaviour of the derivative of the flow velocity at each point on the critical
curve. If according to this equation, ${\partial \dot{\phi}_i \over \partial \phi_j}$ becomes singular at any
point other than the initial one, then $S_{,\phi_i \phi_j}$ becomes unbounded here. 

Consider next the case of the boundary condition such that at $\rho = \rho_1$ and $\phi_i(\rho_1)$ is
unspecified. Then we have seen that at the line $\rho = \rho_1$, $S$ and its partial derivatives are zero.
Then for this choice of boundary condition, the initial conditions on (\ref{umaster}) are such that at $\rho =
\rho_1$ :
\begin{equation}
{\partial \dot{\phi}_k \over \partial \phi_j}F_{,\dot{\phi_i} \dot{\phi}_k} + F_{,\phi_i \dot{\phi}_j}~=~0.
\end{equation}
Thus we now have a criterion to determine the applicability of the fundamental equation (\ref{master}). 

The above considerations are next applied to the example of a single scalar field.
For a single scalar field, the matter lagrangian part of the $AdS_{d+1}/CFT_{d}$ example discussed in the
previous section  becomes :
\begin{equation}
{\cal L}_M~=~\sqrt{g} \left( V(\phi) + \frac{1}{2}\nabla_{\mu} \phi \nabla^{\mu} \phi \right),
\label{1scalar}
\end{equation}
and for this case Eq.(\ref{umaster}) for the scalar may be written as :
\begin{equation}
\left({d \over d\rho} \left( F_{,\phi \dot{\phi}} + F_{,\dot{\phi}\dot{\phi}}{\partial \dot{\phi}
\over \partial \phi} \right) \right)_{\dot{\phi}} - F_{,\phi \phi} + F_{,\dot{\phi}\dot{\phi}}
\left({\partial \dot{\phi} \over \partial \phi}\right)^2~=~0.
\label{umaster2}
\end{equation}
In this equation we make a change of variables :
\begin{equation}
{\partial \dot{\phi} \over \partial \phi}~=~{\dot{W} \over W},
\label{changev}
\end{equation}
so that it now becomes :
\begin{equation}
F_{,\dot{\phi}\dot{\phi}} \ddot{W} + {d \over d\rho} F_{,\dot{\phi} \dot{\phi}} \dot{W} + 
\left( {d \over d\rho} F_{,\phi \dot{\phi}} - F_{,\phi \phi} \right) W~=~0.
\label{umaster3}
\end{equation}
It is easy to see that the question of the unboundedness of the LHS of Eq.(\ref{changev}) now translates to
the existense of zeros of $W$. Thus with the choice of boundary conditions such that at $\rho = \rho_1$,
$\phi = \phi_1$, we know that $W$ has a zero at the initial point and to check if the second derivatives of
$S$ are unbounded, we must  study the conditions under which Eq.(\ref{umaster3}) implies  other zeros of
$W$. In fact, using Eq.(\ref{1scalar}), Eq.(\ref{umaster3}) may be written as ,
\begin{equation}
\ddot{W} + K(\rho) \dot{W} - u(\phi) W~=~0,
\label{umaster4}
\end{equation}
where, $K$ is the extrinsic curvature (we are using the gauge choice of Eq.(\ref{gauge})), and $u(\phi) =
V_{,\phi \phi}$. It is clear from Eq.(\ref{umaster4}) that if $u(\phi) \geq 0$, $W$ has at most one zero. 
More generally, let us make a furthur change of variables from $W$ to $Y$ : 
\begin{equation}
W~=~ e^{\int \lambda d \rho} Y.
\end{equation}
Then the zeros of $W$ are unchanged for finite $\lambda$, and Eq.(\ref{umaster4}) becomes:
\begin{equation}
\ddot{Y} + (K + 2\lambda) \dot{Y} + (-u + \lambda^2 + \dot{\lambda} + \lambda K) Y~=~0.
\end{equation}
We now choose $-2 \lambda=K$ and the above becomes :
\begin{equation}
\ddot{Y} - (u + {K^2 \over 4} + {1 \over 2}\dot{K}) Y~=~0.
\label{finalex}
\end{equation}
Thus, quite generally, we see that if $u$ is negative, but $K$ is large enough and $\dot{K}$ is not too large (and
negative),  such that,
\begin{equation}
(u + {K^2 \over 4} + {1 \over 2}\dot{K}) \geq 0,
\label{condition}
\end{equation}
then, $Y$ and hence $W$ will have only one zero. If not, then the second derivatives of $S$ become unbounded
and  Eq.(\ref{master}) is no longer applicable - the quantum mechanical boundary theory cannot just be described
by a classical theory in the bulk (tree level supergravity in the case of AdS) .  More specifically, in order for the 
classical description of the bulk to be equivalent to the (quantum) boundary theory, at each point along the critical
trajectory the trace of the extrinsic curvature of the foliation at that point must be related to the parameters
of the classical theory  through the condition (\ref{condition}). One way to satisfy the condition for the AdS case 
is that the critical trajectory must also minimize the potential $V$ as well. It  appears difficult to
find another possibility for the AdS case since in this case $K \sim 1/r$ and for holography $r$ is large
\cite{suskindreview}. Note that in Eqs.(\ref{finalex}) and (\ref{condition}) we have retained a term
$\dot{K}$ even though it vanishes for the specific case of constant curvature.We do so because in more general 
situations such a contribution will be present  for  either sign of $u$.  This is a particularly interesting term 
and becomes important when $K$  is changing very rapidly at the boundary. Eq.(\ref{condition}) then tells us that when
this occurs a holographic description  becomes questionable. This is an important conclusion of our analysis. 
Furthur study of this "non-equilibrium" type situation is, however, beyond the scope of this paper.

\section{Domain Walls and Radial Flow} 
Until now we have discussed the evolution in $\rho$ in a $d+1$
dimensional space-time with no hypersurfaces embedded in it. Consider  next a Domain wall in this
$d+1$ dimensional space-time.  It has $d-1$ spatial dimensions (a $d-1$ brane) and partitions the space-time
into different domains.  The presence of such domain walls can change the physical properties of the theory, in
particular, its  spectrum.  For example, if in the original theory, the range of $\rho$  was infinite, the domain
wall will change this, perhaps making it  semi-infinite.  Additional normalizeable fluctuations of certain
fields (for example the metric in theories with gravity) will therefore make thier appearance where
previously (in the absence of the wall)  they were absent. In the following we will limit our considerations to
a single such domain wall and generalizations can be straightforwardly accomodated.

At domain wall junctions, the flow velocities can become discontinuous. For example, in theories with    
gravity it is common in such situations that the metric $g_{\mu \nu}$ $\in$ $C^0$ \cite{israel}. We are 
interested in
finding    out how the partial derivatives of $S$ behave as we cross the domain wall junction. Consider:
\begin{equation}
{d \over d\rho} \left( {\delta S \over \delta \phi_i} \right)~=~{\partial \over \partial \rho}
{\delta S \over \delta \phi_i} +
{\delta^2 S \over \delta \phi_i \delta \phi_k}\dot{\phi}_k
\end{equation}
Comparing this with Eq.(\ref{euler2}) we get:
\begin{equation}
{d \over d\rho} \left( {\delta S \over \delta \phi_i} \right)~=~F_{,\phi_i}.
\label{junc1}
\end{equation}
If we denote by $Q^+$ and $Q^-$ the values of any quantity on the left and right respectively of the domain wall,
then :
\begin{equation}
S_{,\phi_i}^+ - S_{,\phi_i}^-~=~\int_{0^-}^{0^+} d\rho F_{,\phi_i}.
\label{junc2}
\end{equation}
SImilarly, we have for the partial derivative w.r.t. $\rho$ :
\begin{equation}
{d \over d\rho}\left({\partial S \over \partial \rho}\right)~=~{\partial^2 S \over \partial \rho^2} +
{\partial \over \partial \rho}\left( {\delta S \over \delta \phi_k} \right)\dot{\phi}_k
\label{junc3}
\end{equation}
Then from Eq.(\ref{hj}) we have :
\begin{equation}
F_{,\rho} + F_{,\dot{\phi}_k}{\partial \dot{\phi}_k \over \partial \rho}~=~{\partial^2 S \over \partial \rho^2} +
 {\partial \over \partial \rho}\left( {\delta S \over \delta \phi_k} \right) \dot{\phi}_k + 
S_{,\phi_k}{\partial \dot{\phi}_k \over \partial \rho}
\end{equation}
Combining this with Eq.(\ref{junc3}) we get:
\begin{eqnarray}
{d \over d\rho}\left({\delta S \over \delta \rho}\right)~=~F_{,\rho} : \\
S_{,\rho}^+ - S_{,\rho}^-~=~\int_{0^-}^{0^+} d\rho F_{,\rho}.
\label{junc4}
\end{eqnarray}
In deriving the above Eqs.(\ref{junc2}, \ref{junc4}) we have implicity assumed the conditions under which 
the flow velocities may be discontinuous across the domain wall. A condition, obtained from an
examination of Eq.(\ref{hj2}) is: the change in
$F_{,\dot{\phi}_i}$ across the wall is finite (i.e., $F_{,\dot{\phi}_i} \in C$), which can be used together with
Eq.(\ref{hj}) to obtain others.
Eqs.(\ref{hj}, \ref{hj1} and \ref{hj2}) and Eq.(\ref{junc2}) can be used to find the equations of motion of the various
fields at the domain wall.  The functional form of $S$ is determined from the first of these and the second gives the
equation of motion at the domain wall. We will next apply the above ideas to the case of a domain wall in
$AdS_{d+1}$, and the related Randall-Sundrum type scenario.

The Randall-Sundrum model may be described by the action (with Euclidean signature) :
\begin{equation}
-\int_{d+1}\sqrt{g^{d+1}}(R^{d+1} + 2\Lambda) -2\int_{d}\sqrt{g}K + T \int_d\sqrt{g},
\label{rs}
\end{equation}
where matter contributions may be added-we are neglecting these for simplicity;
and $T$ denotes the brane tension. Randall and Sundrum \cite{sundrum} were able to show (actually for
$d=4$) that for a certain value of $T$, gravity may be "trapped" on the $d-1$ brane with an
effective Newton constant of ${d-2 \over 2r}$. Using the picture of radial flow for a bulk
AdS theory developed in previous sections we will see how this may be understood simply in terms of a 
correspondence with a CFT at some finite value of $\rho$ (i.e., away from the boundary of AdS) where 
the domain wall is situated. Indeed, consider the AdS bulk action of Eq.(\ref{adslag}) without the scalars, 
and to it let us add a brane action of the form,
\begin{equation}
I_{DW}~=~T\int_d d^dx \sqrt{g}~=~T\int_{d+1}d^dx d\rho \sqrt{g}\delta(\rho);
\end{equation}
where as discussed before, the domain wall is located at $\rho=0$. We will next use the junction
conditions derived in Eq.(\ref{junc2}). Since only the domain wall action contributes to the
RHS, for the present case this equation reduces to
\begin{equation}
S_{,g_{\mu\nu}}^+ - S_{,g_{\mu\nu}}^-~=~T \sqrt{g}\left( \frac{1}{2} g^{\mu\nu} \right).
\label{adsjunc2}
\end{equation}
In the previous sections we have obtained the form of (the local part of) the effective action $S$
at an arbitrary foliation of AdS with the result:
\begin{equation}
S_{loc}~=~\int_d d^dx \sqrt{g}( U + \Phi_1 R + \Phi_2 R^2 + \Phi_3 R_{\mu\nu}R^{\mu\nu} +
\Phi_4 R_{\mu\nu\alpha\beta}R^{\mu\nu\alpha\beta} + ........,
\end{equation}
with the following values for the various parameters :
\begin{eqnarray}
U^2~=~ {4(d-1) \over d}2\Lambda, \\
\Phi_1~=~ {2(d-1) \over d-2} \frac{1}{U}, \\
\Phi_2~=~  -\frac{1}{2}{d \over d-4}{\Phi_1^2 \over U}, \\
\Phi_3~=~ 2{d-1 \over d-4}{\Phi_1^2 \over U}, \\
\Phi_4~=~0.
\end{eqnarray}
Returning to Eq.(\ref{adsjunc2}), we now make the simplifying assumption that the entire AdS
space has a $Z_2$ symmetry so that $S_{,g_{\mu\nu}}^+ = - S_{,g_{\mu\nu}}^- =
S_{,g_{\mu\nu}}$. Then,
\begin{equation}
 2S_{,g_{\mu\nu}}~=~T \sqrt{g}\left( \frac{1}{2} g^{\mu\nu} \right).
\end{equation}
From this it is easy to obtain a variety of conditions; first, we see that in order for the linear
terms to cancel (fine tuning of the $d$ dimensional cosmological constant) we must have 
\begin{equation}
T~=~2U~=~-4{d-1 \over r},
\end{equation}
which is just the Randall-Sundrum condition for the brane tension. Secondly, we see that
induced gravity on the brane has various contributions , with the leading and next to leading
corrections calculated above. From the coefficient $\Phi_1$ we notice the Randall-Sundrum
relation for the $d$ dimensional gravitational constant, i.e., 
\begin{equation}
G_d~=~{d-2 \over 2r}.
\end{equation}
The coefficients $\Phi_i, i=2,3,4$, give the corrections to Newton's law. These give the same
results as before (see ref.\cite{sundrum} and the last reference in \cite{connection}). From the
above we see that in general the induced gravity on the  brane is the same as for any foliation of 
AdS at some value of $\rho$ coincident with the brane position.  The generating function for the 
conjectured dual conformal field theory is related to the non-local part of this  effective action $S$. 
This conformal field theory on the brane, however, will in general, be different than the one at the 
boundary of AdS. 

\section{Conclusions and Outlook}
We have developed a variational method for studying the evolution of bulk fields with the radial
coordinate in holographic theories. The method appears quite general and is coordinate independant.
It is worth pointing out that the entire formalism may be used to derive and to study the exact
renormalization  group equations in $d$-dimensions with a fictitious $d+1$th coordinate playing the role of the
renormalization group parameter. The resulting equations will be the same as in section 2 which in turn       
differ from the Polchinski \cite{polchinski} renormalization group equations in that they do not contain terms 
with the second partial derivatives of the effective action. The conditions when this is a good approximation have 
been studied in  section 4 where certain differential equations were derived. If the solutions of these equations 
become unbounded then the terms with the second partial derivatives of $S$ must be included. The method of 
including these will be discussed in a subsequent publication and is related to the stochastic quantization method.

The variational approach has allowed us to look at the boundary conditions for the flow equations in a novel way
and to formulate the criteria for thier validity in a quantitative manner. It would be very instructive to
determine the conditions for unbounded solutions to Eq.(\ref{umaster}) for the AdS case with many scalar 
fields in a similar manner as for a single scalar. Bulk space-times other than of the AdS type also 
need to be investigated in detail. Again the  developments in this paper should be useful for this. We have 
also studied more general situations which include domain walls in bulk space-time. Methods were developed 
by combining the flow equations with the generalized junction conditions to study aspects of gravity in these 
brane-world scenarios. Again we are  trying to develop other more realistic 
scenarios than discussed in the paper.

Another advantageous aspect of the variational approach which we have not utilized in this paper involves
a systematic study of symmetry properties. Indeed, it is straightforward to obtain Ward-like
identities for the effective action. It would be interesting to see what information about the boundary field theory
one may obtain when these are combined with the flow equations. Work in these directions is in progress. 

\section{Acknowledgements} This work was supported in part by the US Department of Energy.

\end{document}